\documentclass[12pt,thmsa]{article}
\usepackage{amssymb}

\usepackage{graphicx}

\input{tcilatex}
\begin{document}

\begin{center}
\textbf{\Large Aspects of a Two-Level Atom in Squeezed Displaced Fock States
 }

\textbf{\ }

{\textbf{A.-S. F. Obada$^1$, G. M. Abdal-Kader$^1$  and Mahmoud Abdel-Aty$^{2,}$\thanks{E-mail: abdelaty@hotmail.com}}
} 
~

{\small $^1$Mathematics Department, Faculty of Science, Al-Azhar University,  Cairo, Egypt\\
$^2$Mathematics Department, Faculty of Science, South Valley University,
Sohag, Egypt
 }

\end{center}

\textbf{Abstract:}This paper presents some results on some aspects of the two-level atom interacting with a single-mode with the privileged field
mode being in the squeezed displaced Fock state (SDFS).
The exact results are employed to perform a careful investigation
of the temporal evolution of the atomic inversion, entropy and phase distribution.  
It is shown that  the interference between component states leads to
non-classical oscillations in the photon number distribution.
At mid revival time the field is almost in the pure state.
 We have briefly discussed  the evolution of the Q function of the cavity field.
The connection between the field entropy and the
collapses and revivals of the atomic inversion has been established.
We find that the phase probability distribution of the field reflect the
collapses and revivals of the level occupation probabilities in most
situations. The interaction brings about the symmetrical splitting of the phase probability distribution.
The general conclusions reached are illustrated by numerical results.


~

\begin{center}
To be published in Physics Scripta
\end{center}

\newpage
\section{ Introduction}

The entropy of a radiation field is one of the canonical problems 
of statistical physics and has 
attracted much attention in the past. 
In recent years much attention has been focused on the properties of the
entanglement between
the field and the atom and in particular the entropy
of the system
[1-11]. The authors in [2-4]
have shown that
entropy is a very useful operational measure of the purity of the
quantum
state, which automatically includes all moments of the density operator.
The time evolution of the field (atomic) entropy reflects the time
evolution
of the degree of entanglement between the atom and the field.
The higher the entropy, the greater the entanglement.

The concept of the photon in the quantum theory of a radiation field
has been built on the number (Fock) state $|n\rangle $.
However, another important state is the coherent state
which is a linear superposition of all $|n\rangle $ states with
coefficients chosen such that the photon number  distribution is
Poissonian. It may be  defined by
the action of a displacement operator $D(\alpha)$ on the vacuum state.
This state  has been
extensitevly  studied [12-14].  On the other hand
the squeezed state is
one of the non-classical states of the electromagnetic field in which
certain observables exhibit fluctuations less than in the vacuum
state [15-17]. It is defined by the action of the squeeze operator $S(z)$ on
the coherent state [15-17].   Squeezed displaced Fock states (SDFS)
 have been introduced and different aspects of these
states such as   squeezing  and  photon
statistics have been investigated [18-25].
These states generalize two-photon coherent states [17]
(squeezed coherent states), squeezed number states [18], and
displaced Fock states [26-28].
 They exhibit both number
 and  quadrature squeezing.
 Recently
the creation of nonclassical states of motion of
a trapped ion such as Fock states, coherent states, squeezed
states and  Schr{\"o}dinger-cat states have been reported [29-31]. In
these experiments an ion is
laser-cooled in a Paul trap to the ground harmonic state. Then the
 atom is put into various quantum states of motion by applications of
optical and electric fields.   That moved the study  of these
states from
the academic realm to the world of experimentation.
This motivated us to study the interaction  of these states with  a two-level atom.

A stochastic formulation of quantum mechanics involves, basically,
two interrelated problems. These are the determination of the
probability functions of the density operator, $ {\rho} $, and the
establishment of the proper correspondence between quantum-mechanical
observable, and ordinary functions in phase-space. Attempts
in this direction ran into the difficulty of dealing with
quasidistributions,[32-33]. The three types of quasiprobability
distribution functions P, Q and W (for normal, antinormal and symmetric
ordering respectively) are very important in quantum optics [32-33].

Recently Pegg and Barnett [34, 35] have  introduced a new hermitian phase
formalism which successfully overcomes the troubles inherent in the
Susskind-Glogower phase formalism and enables one to study finer
detials of the phase properties of quantum fields. Such quantities as
expectation values and variances of the hermitian phase operators or
phase distribution functions are now available for investigation
[36, 37]. One of our interests is to investigate the phase properties
here.

The material of this paper is arranged as follows:
In section 2, we review a few
concepts of  squeezed displaced Fock states (SDFS's).
In section 3 we introduce the model and write the
expressions for the
final state vector at any time $t> 0 $.
We discuss the field entropy
in section 3.1.
By  numerical computations, we examine the influence of the
SDFS's on the field entropy evolution and entanglement
of the atom and the
field. We analyse oscillations in the photon number distribution of the cavity
field
in section 3.2.
Section 3.3  discuss the phase probability distribution through the framework
of
Pegg-Barnett's  definition of the Hermitian phase operator.
We present the evolution of the Q function
for the SDFS's in section 3.4.
Finally, summary and remarks are presented.

\section{Squeezed displaced Fock states (SDFS's)}

 The  SDFS's are generated from the number state $|m\rangle$ as shown below.
These states have been studied extensively in literature,
because of their interesting nonclassical properties and expected
prospective applications in optical communication and interferometery
[14]. The  SDFS is defined by [18-25]
\begin{eqnarray}
 {\vert} {\alpha_0},z,m{\rangle} =  D({\alpha_0}) S(z) {\vert}m{\rangle}
\end{eqnarray}
where the displacement operator $ D({\alpha_0})$ ,( with $ \alpha_0 $
a complex parameter that represents the magnitude and angle of the
displacement) [12-13],
and the squeeze operator
$ S(z) $ are given by [14-17]
\begin{eqnarray}
D({\alpha_0})&=&\exp({\alpha_0}{a^+}- {{\alpha_0}^*}a),
\nonumber
\\
S(z)&=&\exp[{\frac{z^*}{2}a^2}-{\frac{z}{2}{{a^+}^2}}],
\end{eqnarray}
where $z=r e^{i \phi} $ and $r$ is  known as the  squeeze parameter
and $\phi$ indecates the  direction of the squeezeing, 
with $ a$  $ (a^+) $ are the annihilation (creation) operators of the
field.

It is easy to calculate the average value of the number
operator, ${a^+} a$, in the state $ {\vert} {\alpha_0},z,m{\rangle} $,
by using the above relations, thus
\begin{eqnarray}
{\langle}{a^+}a{\rangle}=({\vert}{\mu}{\vert}^2+{\vert}{\nu}{\vert}^2)m
+{\vert}{
\nu}{\vert}^2+{\vert}
{{\alpha_0}}{\vert}^2,
\end{eqnarray}
where    $\mu=\cosh{r}$,
and $\nu=\exp(i\phi)\sinh{r} $.

The photon number distributions $ P_n $ for
${\vert}{\alpha_0},z,m{\rangle} $ is equal to the square of the absolute
value of the matrix elements $ {\langle}n{\vert} {\alpha},z,m{\rangle} $.
The analytical expression for $ {\langle}n{\vert}{\alpha},z,m{\rangle}
$ is  given by [19],
\begin{eqnarray}
{\langle}n{\mid}&{\alpha_0},z,m{\rangle}
=({\frac{n!}{{\mu}m!}})^{1/2}
({\frac{\nu}{2\mu}})^{n/2}
\exp(-\frac{{\mid}{\bar \alpha_0}{\mid}^2}{2}
+\frac{{\nu}^*}{2{\mu}}{\bar
\alpha_0}^2)
\nonumber\\&
\times
{\sum_{i=0}^{min(n,m)}} {(^{m}_{i})} 
\frac {({\frac{2}{\mu
\nu})}^{i/2}}{(n-i)!}
\lbrack(\frac{-{\nu}^*}{2\mu})^{\frac{m-i}{2}}
\nonumber\\&
\times
\rbrack {H_{n-i}\lbrack{\frac{\bar \alpha_0}{(2{\nu}{\mu})^{1/2}}\rbrack}
{H_{m-i}\lbrack{\frac{-{\alpha_0}^*}{(-2{\nu^*}{\mu})^{1/2}}
\rbrack}}},
\end{eqnarray}
where $ \bar \alpha_0=\mu\alpha_0 +\nu {\alpha_0}^*$, and  $H_n(x)$
stands for the  Hermite function of order $n$ [38].
Therefore, the photon number distribution  $ {P_n} $ is given by
\begin{eqnarray}
P_n={{\vert}{\langle}{n}{\vert}{\alpha_0},z,m{\rangle}{\vert}}^2 ,
\end{eqnarray}

The scalar product of two different SDFS states
$\langle \alpha_1,z_1,m_1|\alpha_2,z_2,m_2\rangle  $ is very useful in the
representation with SDFS basis. We can obtain it with the help
of the completeness of coherent states as 
\begin{eqnarray}
\langle \alpha_1,z_1,m_1|\alpha_2,z_2,m_2\rangle  =\frac{1}{\pi} \int
\langle \alpha_1,z_1,m_1|\beta\rangle  \langle \beta| \alpha_2,z_2,m_2\rangle   \thinspace d^2\beta
\nonumber
\end{eqnarray}
By using equation (4),

\begin{eqnarray}
\langle&&\alpha_1,z_1,m_1|\alpha_2,z_2,m_2\rangle=
 \frac{1}{\sqrt{m_1!m_2!
\mu_1\mu_2 K}} (-\frac{\nu_2^*}{2\mu_2})^{m_2/2}
\nonumber
\\&&
\times
(-\frac{\nu_1}{2\mu_1})^{m_1/2} 
\exp\biggl[ -\frac{|\alpha_2|^2}{2}  
-\frac{|\alpha_1|^2}{2}\biggr] 
\nonumber
\\&&
\times\exp\biggl[\frac{1}{\mu_2\mu_1-\nu_2\nu_1^*}
\alpha_1^* \alpha_2 
 -\frac{\nu_2 \mu_1 -\mu_2 \nu_1}{ 2 (\mu_2\mu_1-\nu_2\nu_1^*)}
{\alpha_1^*}^2
\nonumber
\\&&
+\frac{(\nu_2 \mu_1 -\mu_2 \nu_1)^*}{2(\mu_2\mu_1-\nu_2\nu_1^*)}
 {\alpha_2}^2 \biggr]
\sum_{s=0}^{[m_2/2]}\sum_{\acute s=0}^{[m_1/2]}
\sum_{r=0}^{min(m_2-2s,m_1-2\acute s)}
\nonumber
\\&&
\times
 \frac{m_1!m_2!(-1)^{s+\acute s} }
{s! \acute s! r!} 
\biggl[ \frac{2}{(-2\mu_2 \nu_2^*)^{1/2}}\biggr]^{m_2-2s} 
\nonumber
\\&&
\times
\biggl[\frac{2}{(-2\mu_1 \nu_1)^{1/2}}\biggr]^{m_1-2\acute s} 
\frac{(1/K)^r}{(m_2-2s-r)!(m_1-2\acute s-r)!}
\nonumber
\\&& 
\biggl(\sqrt{\frac{\nu_2}{2K \mu_2}}\biggr)^{m_1-2\acute s-r}
H_{m_1-2\acute s-r} 
\nonumber
\\&&
\times
\biggl[
\frac{(\alpha_{02}-\alpha_{01})-\frac{\nu_2}{\mu_2}(\alpha_{01}^* 
-\alpha_{02}^*)} {2\sqrt{\frac{K\nu_2 }{2\mu_2}}} \biggr]
\biggl(\sqrt{\frac{\nu_1^*}{2K \mu_1}}\biggr)^{m_2-2s-r}
\nonumber
\\&&
\times
H_{m_2-2 s-r}\biggl[
\frac{(\alpha_{01}^*-\alpha_{02}^*)-\frac{\nu_1^*}{\mu_1}(\alpha_{02}
 -\alpha_{01})} {2\sqrt{\frac{K\nu_1^* }{2\mu_1}}}\biggr]
\nonumber
\end{eqnarray}
where
$
K={\mu_2\mu_1-\nu_2\nu_1^*}/{\mu_2\mu_1}
$
and
$
\alpha_{01}={\mu_1}\alpha_1 -\nu_1\alpha_1^* , $ $  
\alpha_{02}={\mu_2}\alpha_2 -\nu_2\alpha_2^*
$. An equivalent result (however not having the Hermite polynomials in
the summand) has been given recently [24].
When $m_1=m_2=0$  the above equation gives the result in [16], for the
squeezed state. But when $ \nu_1=\nu_2=0 $ and $\mu_1=\mu_2=1$ then
this equation becomes 
\begin{eqnarray}
&&\langle \alpha_1,m_1|\alpha_2,m_2\rangle 
=\frac{\langle \alpha_1|\alpha_2\rangle}{\sqrt{m_1
m_2}} \sum_{r=0}^{min(m_1,m_2)} 
\nonumber
\\&&
\times
\frac{m_2! m_1!}{r! (m_2-r)!
(m_1-r)!}
(\alpha_1^* - \alpha_2^*)^{m_2-r} 
(\alpha_2-\alpha_1)^{m_1-r}
\nonumber
\end{eqnarray}
which is the result of, [28,18], for the displaced Fock state. Other
special cases follow in a straightforward way.

\section{The quantum dynamics }

 We consider the Hamiltonian for the one-photon Jaynes-Cummings
model. It describes the interaction of a
single-mode quantized  field with a two-level atom via
a one-photon process. The  Hamiltonian
 of the system in the
rotating-wave approximation is written as

\begin{eqnarray}
{\hat H}=\frac{1}{2} \omega_\circ \sigma_z+\omega {\hat a}^\dagger
 {\hat a}
+\lambda ({\hat a}
\sigma_++{\hat a}^{\dagger  }\sigma_-),
\end{eqnarray}

where $ {\hat a}^\dagger $ $({\hat a})$ is
 the creation (annihilation)
operators for the photon of frequency $ \omega, $ and $\lambda  $
\quad  describes  the coupling to the atomic system. The two level
atom with transition frequency  $ \omega_\circ  $ is described by the
 Pauli raising and ( lowering ) operators $ \sigma_+, (\sigma_- )$
and the inversion operator $ \sigma_z $.
with the detuning parameter $ \Delta=\omega-\omega_\circ $.

The initial state of the total atom-field system can be written as

\begin{eqnarray}
\mid \psi(0)>=\psi(0)>_f\otimes\mid\psi(0)>_a
 =\sum_{n=0}
q_n\mid
n,e>,
\end{eqnarray}
means that the atom starts in its excited state,
the field is assumed to be initially
in the squeezed displaced Fock
states(SDFS)
$ q_n= <n\mid \alpha_0,z, m>
$ given by Eq. (4). The solution of the Schr{\"o}dinger equation in the interaction picture
i.e  the wave
function of the system at  any time $ t>0 $\quad is given by

\begin{eqnarray}
\mid \psi(t)>=\sum_{n=0}^\infty \biggl(A_{n}(t)\mid n,e>+B_{n}(t)\mid
n+1,g>\biggr),
\end{eqnarray}

where the coefficients $ A_{n} $ and $ B_{n} $ are given by the
formulae
\begin{eqnarray}
A_{n}(t)=q_n
\biggl(\cos \lambda t\nu_{n}-\frac{i\Delta}{2\lambda}
\frac{\sin\lambda
t\nu_{n}}{\nu_{n}}\biggr),
\end{eqnarray}

\begin{eqnarray}
B_{n}(t)=-iq_n
\sqrt{n+1}
\frac{\sin\lambda
t\nu_{n}}{\nu_{n}},
\end{eqnarray}

\begin{eqnarray}
 \nu_{n}=\sqrt{
\frac{\Delta^2}{4\lambda^2}
+ n+1 }
\end{eqnarray}
With the wave function $\mid \psi(t)\rangle $ found,
any property related to the atom or the field can be
 calculated.
The reduced density operator  of the field of the system can be written as
$
\rho_f(t)=Tr_{atom}{\mid \psi(t)\rangle\langle\psi(t)\mid}
$,
\begin{eqnarray}
\rho_f(t)&=&\sum_{n,m=0}^\infty  [A_n(t)A^*_m(t)\mid n\rangle\langle m\mid
\nonumber
\\&&
+B_{n}(t)B^*_m(t)
\mid n+1\rangle\langle m+1\mid
\\
&=&\mid C(t)\rangle\langle C(t)\mid +\mid S(t)\rangle\langle S(t)\mid,
\nonumber
\\
\mid C(t)\rangle&=&\sum_{n=0}^\infty 
q_n
\biggl(\cos \lambda t\nu_{n}-\frac{i\Delta}{2\lambda}
\frac{\sin\lambda
t\nu_{n}}{\nu_{n}}\biggr)
\mid n\rangle,
\nonumber
\\
 \mid
S(t)\rangle&=&-i
\sum_{n=0}^\infty q_n
\sqrt{n+1}
\frac{\sin\lambda
t\nu_{n}}{\nu_{n}}\mid n+1\rangle.
 \end{eqnarray}

\subsection{Atomic Inversion}

Using equation (8) we can evaluate the time evolution of the atomic inversion
$W(t)=\langle \psi(t)\mid \sigma_z\mid\psi(t)\rangle $

\begin{eqnarray}
W(t) = \sum_{n=0}^{\infty}\biggl( A_{n}(t) A_{n}^*(t)- B_n(t) B_n^*(t)\biggr),
\end{eqnarray}
with  $P_n=|q_n|^2$ of equation (5)
describing the distribution of the photons in the  state SDFS.
We  start our analysis by taking the squeeze parameter $r=1 $ and for different values of m $(m=0,1,2)$. Since the resulting series cannot be
analytically summed in a closed form, we will evaluate them numerically.
In figure 1a we plot the atomic inversion as a function of scaled time
 $\lambda t $ for the
squeeze parameter $r=1$ and with zero value of  $(m=0)$.

\begin{figure}[htbp]
\begin{center}
\includegraphics[width=8cm]{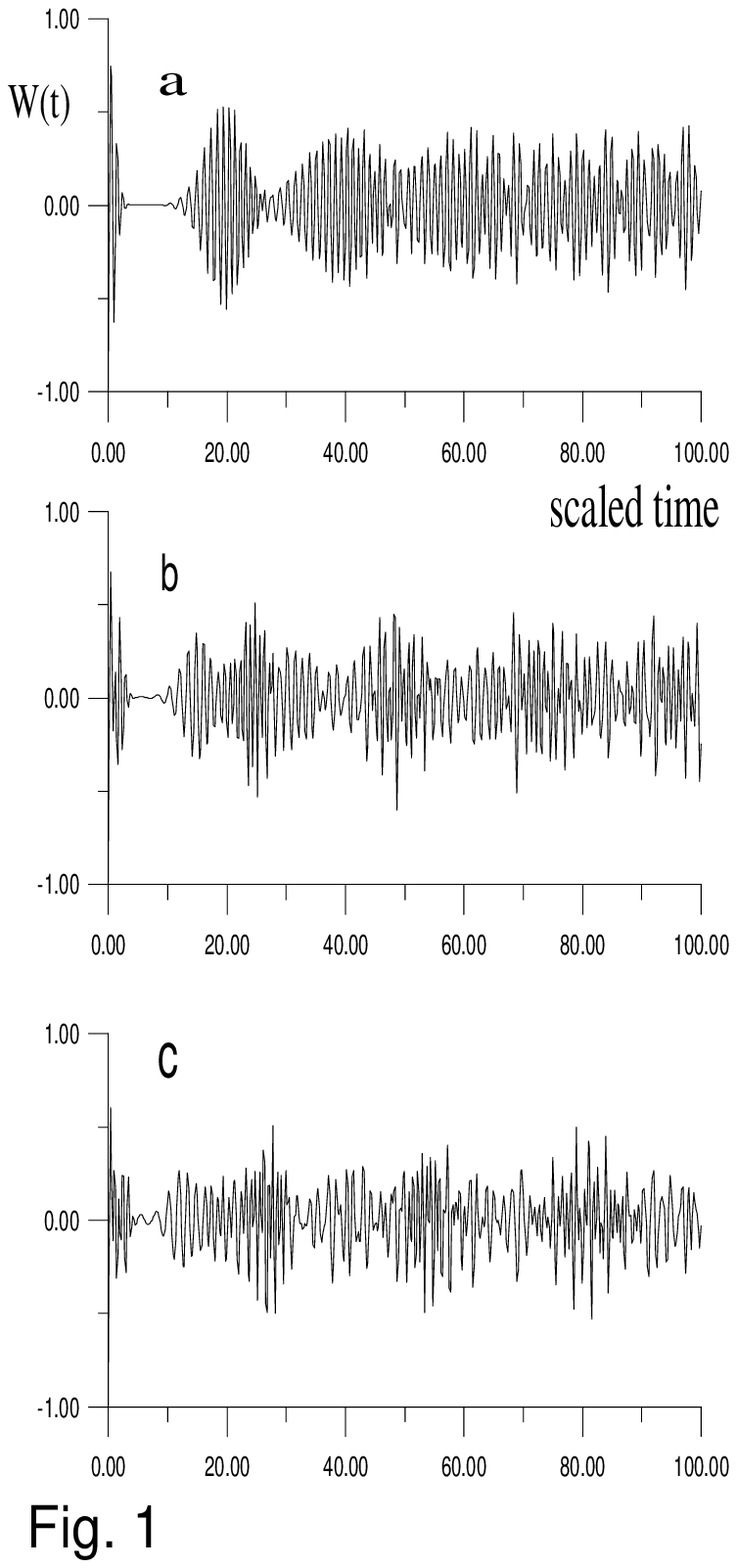}
\end{center}
\caption{ The atomic inversion  
$ W(t)$ for the output field against the scaled time $\lambda t$.
The  values of the parameters are assumed as follows, the squeeze parameter $r=1$, the initial position (3,0), i.e $\alpha_0=3$, and the direction of squeeze is along the x-direction, i.e $\phi=0$. While the number of photons for the input state m takes different values, where (a) $m=0$,  (b)  $m=1$, and  (c)  $m=2$ .
}
\end{figure}

We know that the collapses are caused by the dephasing of the various terms in the sums in equation (14). Thus we can calculate the time in which
the revivals will occur by estimating the time that neighbor terms in the sums will be in phase again (for $n\sim \bar n)$: $T_R(2\lambda\sqrt{\bar
n+1}-2\lambda\sqrt{\bar n})\sim 2\pi.$ This argument is true in the case of
coherent state and can be applied for the squeezed coherent state (i.e, $m=0$), as can be seen in figure 1a, but it cannot be applied for the SDFS.
For nonzero m ($m=1$) figure 1b, not only the amplitude of Rabi oscillations, but also the time average of the inversion is affected. In figure 1c we have increased m to 2, and taken all other parameter as in figure 1a. One observes that for longer time 
the inversion shows small oscillations around zero but in quite irregular manner.

\subsection{Entropy of the cavity  field}

Employing the reduced field density operator given by Eq. (12), we
investigate
 the properties of the entropy. The quantum dynamics described by the
Hamiltonian (6) leads to an entanglement between the field and the
 atom. We use the field entropy as a measurement
 of the degree of entanglement between the field and the atom
of the system under consideration. In order to derive a calculation
formalism of the field entropy, we must obtain the eigenvalues and
eigenstates of the reduced field density operator given by
 Eq. (12). A general
method  to calculate the various field eigenstates in a
 simple way can be found in [4]. By using this method we obtain the
eigenvalues and eigenstates of the reduced density operator,
\begin{eqnarray}
\lambda_f^{\pm}(t)=\langle C(t)\mid C(t)\rangle\pm \exp[\mp\theta]\mid\langle
C(t)\mid S(t)\rangle\mid
\end{eqnarray} \begin{eqnarray}
=\langle S(t)\mid S(t)\rangle\pm \exp[\pm\theta]\mid\langle C(t)\mid
S(t)\rangle\mid,
\end{eqnarray}
\begin{eqnarray}
\mid\psi_f^{\pm}(t)\rangle=\frac{1}{\sqrt{2\lambda_f^{\pm}(t)\cosh
(\theta)}}\{\exp[(i\phi\pm\theta)/2]\mid C(t)\rangle
\end{eqnarray} \begin{eqnarray}
\pm \exp[-(i\phi\pm\theta)/2]\mid S(t)\rangle\},
\end{eqnarray}
where
\begin{eqnarray}
\theta=\sinh^{-1}\biggl(\frac{\langle C(t)\mid C(t)\rangle-\langle S(t)\mid
S(t)\rangle}{2\mid\langle C(t)\mid S(t)\rangle\mid}\biggr).
\end{eqnarray}
We can express the field entropy $S_f(t) $ in terms of the eigenvalue
$\lambda_f^{\pm}(t) $ of the reduced field density operator,
\begin{eqnarray}
S_f(t)=-[\lambda_f^+(t) \ln \lambda_f^+(t)+\lambda_f^-(t) \ln
\lambda_f^-(t)].
\end{eqnarray}

It does not appear possible to express the sums in equation (20) in
closed form, but
for not too large $\bar n $, direct numerical evaluation can be
performed.
On the basis of the analytical solution presented in the previous section,
we shall examine the temporal evolution
of the field entropy.
It should be emphasized that in computing all infinite series for the atomic
 wave function $|\psi(t)\rangle $, we have invoked mathematically sound truncation
criteria. To ensure an excellent  accuracy the behavior of the field entropy
function
$S_f(t) $ has been determined with great precision.
For regions exhibiting strong fluctuation a resolution of $10^3$ point per
unit of time has been employed. The time t has been
scaled; one unit of time is given by the inverse of the coupling
constant $\lambda $.

\begin{figure}[htbp]
\begin{center}
\includegraphics[width=8cm]{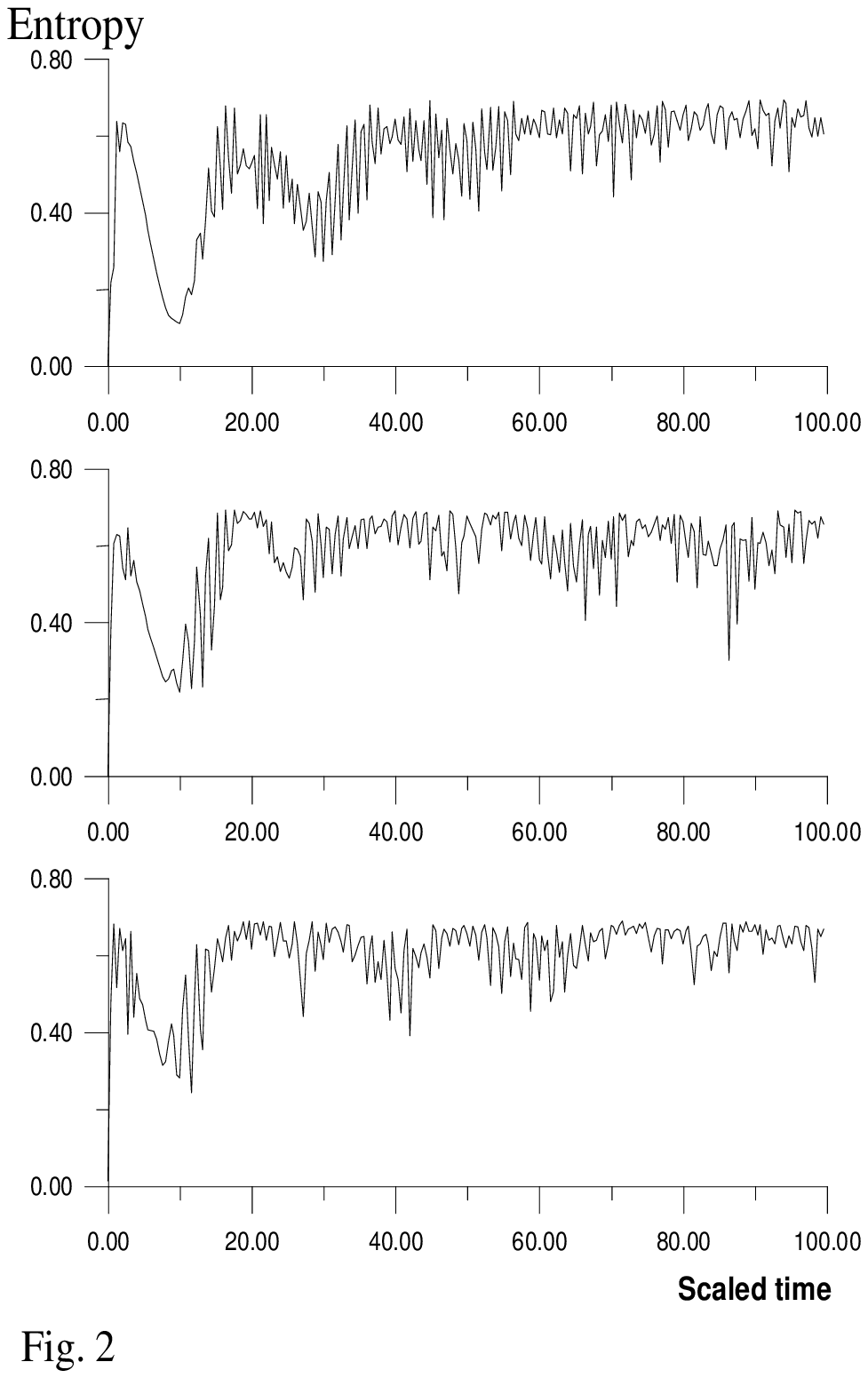}
\end{center}
\caption{Entropy $S_f(t) $ of the output field 
 against  the  scaled time $\lambda t$.
The  values of the parameters are assumed as follows, the squeeze parameter $r=1$, the initial position (3,0), i.e $\alpha_0=3$, and the direction of squeeze is along the x-direction, i.e $\phi=0$. While the number of photons for the input state m takes different values, where (a) $m=0$,  (b)  $m=1$, and  (c)  $m=2$ .
}
\end{figure}

We display the evolution of the  field entropy  for different values
of mean photon number $m$ and the squeeze parameter r is taken to equal $(r=1)$.
 In our computations, we have taken the displacement parameter $\alpha_0=3,  \Delta=0 $.
In the case of an initial field with $r=0 $ (coherent state),
we 
already know [39-40] that the field evolves and (nearly) returns 
to a pure state only at half of the revival time. 
As further analysis showed, at that time the field is indeed
 in a superposition of coherent states. However, 
if we initially prepare the field in a squeezed state, 
as we see in figure 2, ($r=1$) that its entropy reaches a minimum 
at approximately the revival time corresponding to an 
initial coherent state with amplitude $\alpha $. Note that the 
revival time in the case of an initial squeezed state 
is $\lambda t=2\pi(\mid\alpha\mid^2+\sinh^2r)^{1/2} $. 
As we see in figure 2a, in the squeezed state case, 
the value of the entropy is almost the same at the revival 
time and at half of the revival time, and because they 
correspond to minima, the states of the field are less mixed 
at these times. However, near the minima the behavior of the 
entropy is different for the two cases. While the minimum 
is reached in a smooth way for the coherent state, it becomes 
oscillatory for the squeezed. The programme described near $S_f(t) $
has been carried out for several parameter sets, including those 
covered by figure 2.
 It may seem rather surprising to have a pure 
state at a time different from half of the revival time, but
 this is of course due to the nature of the squeezed states.
Also, by increasing the squeezing parameter r, the field entropy
 becomes increasingly irregular and with values characteristic
 of a mixed state, $S_F(t)\simeq 0.7 $. This is in qualitative 
agreement with the fact that the atomic response for the field 
initially prepared in a squeezed vacuum state is very similar to when 
it is prepared in a thermal (mixed) state [41].
To visualize the influence of SDFS in the
field entropy  we set
different values of m $(m=1, 2)$ and the squeeze parameter $r=1$, 
and all the other parameters are the same as in figure 2a.
 The outcome is presented in figure  2b, c. One can
distinguish between two stages of evolution, each of which
has been pictured separately.
We see that 
the amplitude of the oscillations as well as the revival time becomes smaller 
and we have more revivals in the same time, the collapse time decreases.
It is evident that the field and the atom are in the 
entangled  state when m increases further.
In next section we turn our attention to interesting non-classical phenomenon
emerging as a direct consequence of quantum interference between component
states of the field. Namely, we will analyse oscillations in the photon number distribution
of the cavity field at different values of $m $.

\subsection{Photon number distribution}

In direct detection one counts the number of photons in the field
mode of interest. The probabilty for  finding  $n$ photons: at time $t>0$, is given by
\begin{eqnarray}
P(n,t)=| \langle n\mid\psi(t)\rangle |^2 =
 \langle n | \rho_f(t)| n\rangle
\end{eqnarray}

\begin{figure}[htbp]
\begin{center}
\includegraphics[width=8cm]{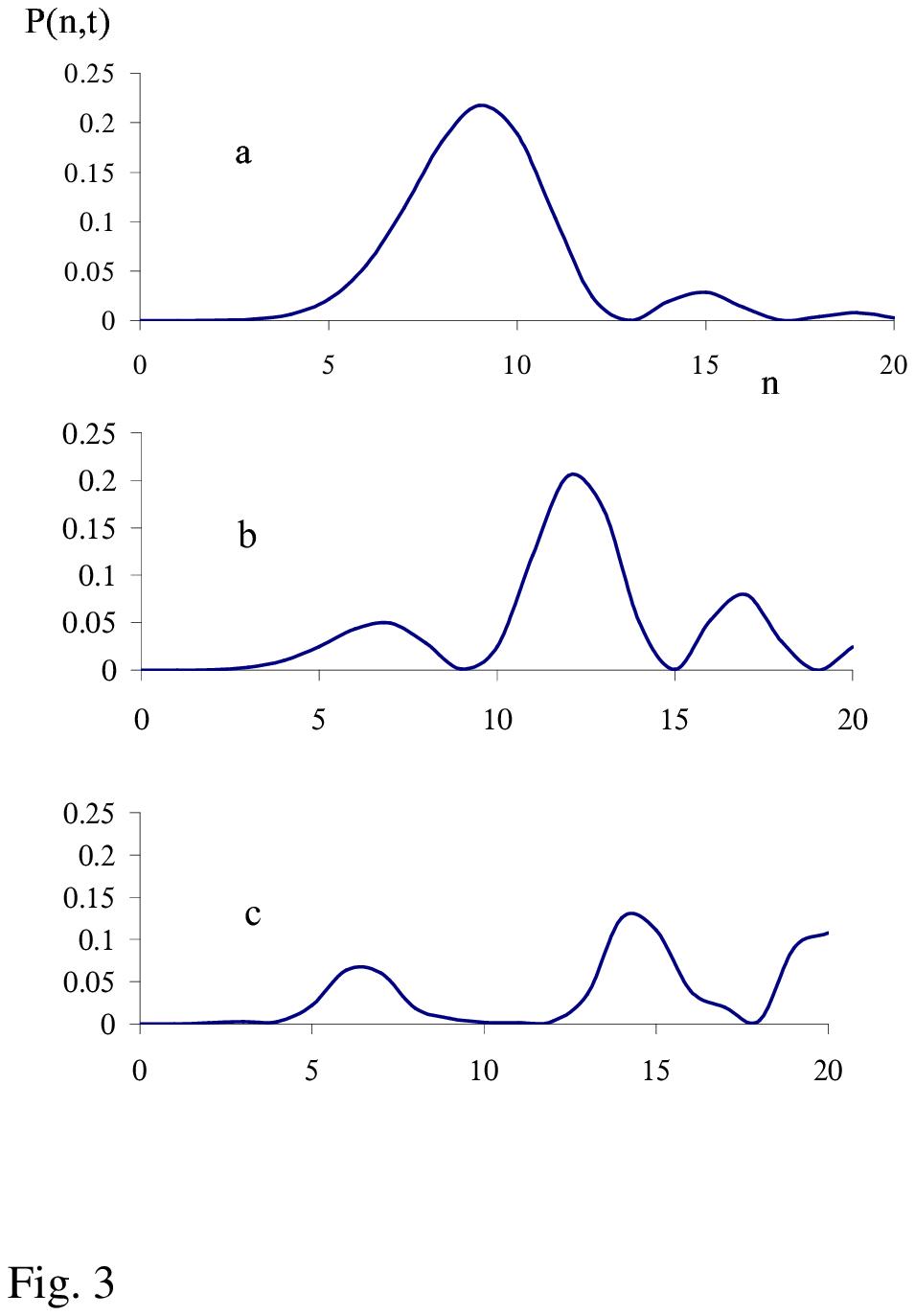}
\end{center}
\caption{Plots for the photon number distribution $P(l,t)$ of the initial field. 
The  values of the parameters are assumed as follows, the squeeze parameter $r=1$, the initial position (0.5,0), i.e $\alpha_0=0.5$, and the direction of squeeze is along the x-direction, i.e $\phi=0$. While the number of photons for the input state m takes different values, where
 (a) $m=0$,  (b)  $m=1$, and  (c)  $m=2$ .
}
\end{figure}

from which we can easily find the photon number distribution of the cavity field
in the one-photon JCM. As the cavity field starts to interact with the atom the
initial  photon number distribution $P_n=\mid q_n\mid^2 $
starts to change. Due to the quantum interference between component states the
oscillations in the cavity field become to be composed of two component
states. Even though the  field entropy is not equal to zero these two component
states partially interfere which results in some oscillations of the photon
number distribution.
In figure 3 we plot the photon-number distribution of the cavity field in the case
of nonzero squeezing parameter $r=1 $ and for different values of m. 
We note that the amplitude of  the photon-number distribution decreases as m increases
see figure 3c (where we have set m=2). 
For $m=0 $( figure 3a),the photon number distribution
resembles a Poissonian distribution. In the large values of m, the 
amplitude of the oscillations is affected. 
As compared with the case $m=0 $, the locations of the maxima
have moved to the right as m increased further. 
 For the entropy the characteristic sequence of minima is there again (figure 2c).

\subsection{Phase distribution}

Recently,  Barnett and  Pegg  defined a Hermitian phase operator in
a finite dimensional state space [34-36].
They used the fact that, in this state space,
one can define phase states rigorously.
The phase operator is then defined as the projection operator on the
particular phase
state multiplied by the corresponding value of the phase. The main
idea of the Pegg-Barnett
formalism consists in evaluation of all expectation values of
physical variables in a finite
dimensional Hilbert space. These give real numbers which depend
parametrically
on the dimension of the Hilbert space. Because a complete description
of the harmonic
oscillator involves an infinite number of states to be taken, a limit
is taken only after the physical results
 are evaluated.  This  leads to
proper limit which
correspond to the results obtainable in ordinary quantum mechanics.
It can be used
for investigation of the phase properties of quantum states of the
single mode of the
electromagnetic field [34-37].

The Pegg-Barnett phase distribution $P(\eta)$
is defined through an infinite sum [34,35]:
\begin{eqnarray}
P(\eta,t)= \frac{1}{2\pi} \sum_{l,j=0}^{\infty} \rho_{lj}(t)
 \exp[i (j-l) (\eta-\eta_0)],
\end{eqnarray}
where the density matrix given by 
\begin{eqnarray}
\rho_{lj}(t)= (A_{l}(t) A^*_{j}(t) + 
B_{l-1}(t)B^*_{j-1}(t) ) 
\end{eqnarray}
the angle $\eta_0$ is the phase reference angle and we take it  to be zero.
The phase distribution can be written as
\begin{eqnarray}
P(\eta,t)= \frac{1}{2\pi}[1+2 Re \sum_{j,l=0;j > l}^{\infty}
\rho_{lj}(t)
 \exp[i (j-l) \eta]
\end{eqnarray}
We have computed
the phase probability distribution function,
 related to a
system of a 2-level atom in interaction with a single mode.
In our computations, we have taken
 the displacement parameter $\alpha_0=3$, and the squeeze parameter  $r=1$.

\begin{figure}[htbp]
\begin{center}
\includegraphics[width=8cm]{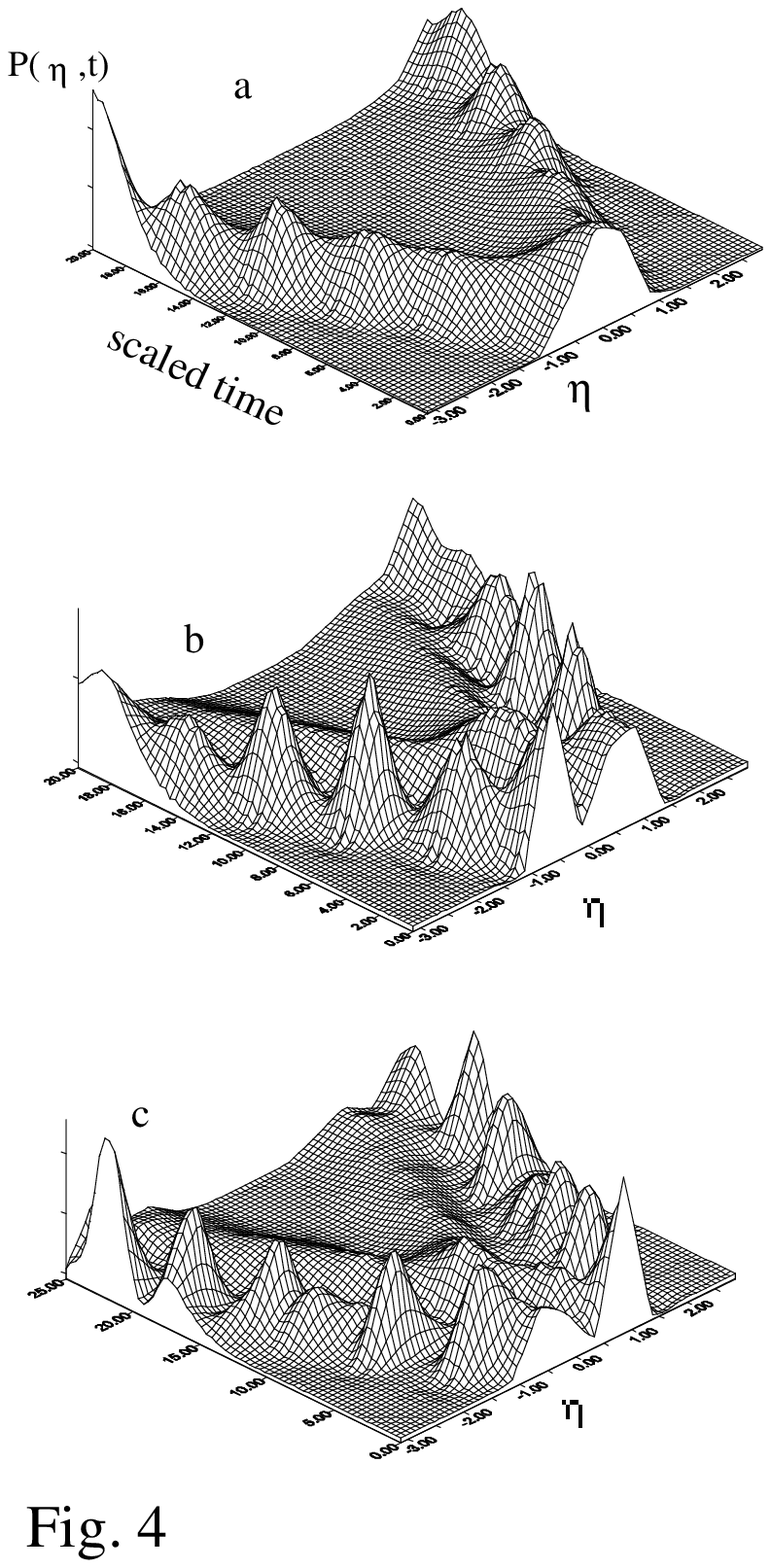}
\end{center}
\caption{ 
Shows the  plot of phase
distribution  $P(\eta,t)$ as a function of the scaled time $\lambda t$ and $\eta$. The  values of the parameters are assumed as follows, the squeeze parameter $r=1$, the initial position (3,0), i.e $\alpha_0=3$, and the direction of squeeze is along the x-direction, i.e $\phi=0$. While the number of photons for the input state m takes different values, where (a) $m=0$,  (b)  $m=1$, and  (c)  $m=2$.
}
\end{figure}

Figure. (4) shows
the time evolution of the phase probability distribution $ P(\eta,t) $ for
 $r=1 $
 and for various values of $m$.
When $m $ equal to zero, it is remarked that $ P(\eta, t)
 $ \quad exhibits symmetric splitting as $\lambda t $
 varies as shown in figure  (4a)
This is the counterrotating behaviour observed earlier [37].
 When $ \lambda t=0 $, $ P(\eta,t) $ has a single-peak structure
 corresponding to the initial coherent state. The peaks are symmetric
 about $ \eta =0 $ so that the mean phase always remains equal
 to zero. The time behaviour of the phase probability distribution
 carries some information about the collapse and revival of Rabi
 oscillations [39].
 When the phase peaks are well separated the Rabi
 oscillations collapse and each time as the peaks meet
( at $\eta=0 $ and/or $\pm \pi $ ) they produce a revival see
 figure  4a. It is further noted that the height of the peak change as time develops in contrast to the case of the coherent input [39]. 
When $ m\ne 0 $,
the situation is completely changed, as we observe from figure  (4b-c).
 It is seen that when
$ m=1,
 $ thee are two peaks, one with  small amplitude
 compared with the other peak, whose rate of shift
 becomes faster when plotted in a phase space as in
 [35]. It is observed that the
symmetry shown in the the case when $m=0 $
for the phase distribution is no longer present
 once the new state is added. The peaks
 are split but the two split peaks move with
different rates. The one with the slower rate
 faces damping while the faster
 peak changes in the amplitude as time develops.

\subsection {Evolution of the Q function }

In the previous section we discussed a particular aspect of the
atomic dynamics (collapses and revivals) in the one-photon model
with the initial field prepared in the  squeezed displaced Fock
states. Now we are going to try to understand better the behaviour
of the system by focusing our attention on the field dynamics by studing the quasi-distribution function.
The first step to be taken is the calculation of the reduced density operator
of the field $\hat \rho_f(t) $ (see equation (12)), we get for the Q function
\begin{eqnarray}
Q(\alpha,t)= \frac{1}{\pi} \langle \alpha | \rho_f(t) |\alpha
\rangle 
\end{eqnarray}
where $|\alpha\rangle$ is a coherent state. More than just a
theoretical curiosity, $Q(\alpha,t)$ can be detected in homodyne
experiments [33] . It
has the form
\begin{eqnarray}
Q(\alpha,t)= \frac{1}{\pi}  \exp(-|\alpha|^2) \sum_{n,m=0}^{\infty}
\rho_{nm}(t)
\frac{ (\alpha^*)^n  (\alpha)^m}{\sqrt{n!m!}}
\end{eqnarray}
The Q function is not only a convenient tool to calculate expectation values
of anti-normally ordered products of operators, but also gives us a new insight
into the mechanism of interaction in the model under consideration.

\begin{figure}[htbp]
\begin{center}
\includegraphics[width=8cm]{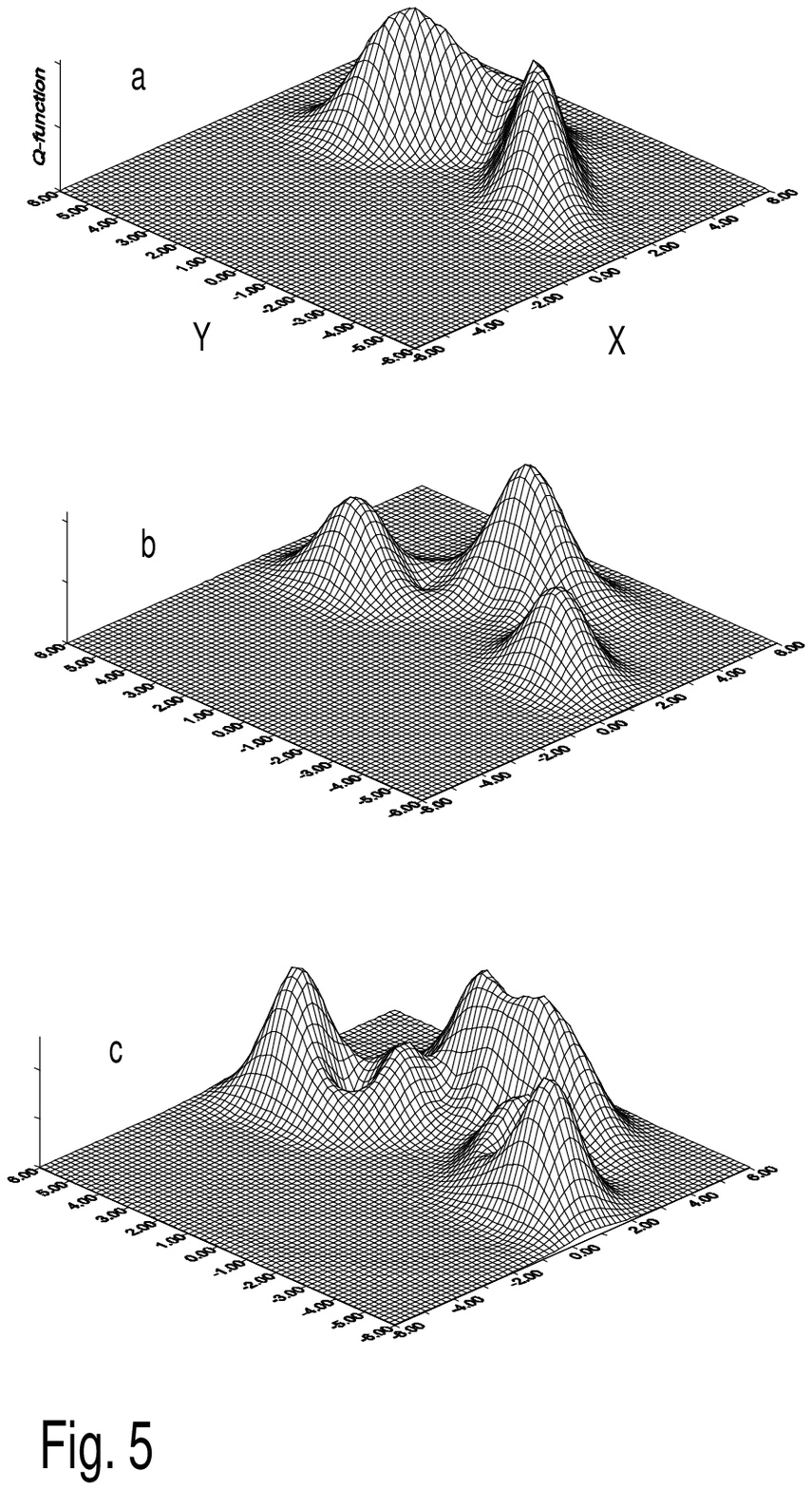}
\end{center}
\caption{ 
The Q  function for the output of the JCM
driven by the SDFS. The parameters are assumed as follows:  The squeeze
parameter $r=1$, $\phi=0$,   $\alpha_0=3$, 
$\Delta=0$, and  $m=1$.   While the interaction 
time  takes the    values: (a) $t=0$;  (b) $t=t_R/2$; (c) $t=t_R$.
 Here $X=Re(\alpha)$ and  $Y=Im(\alpha)$. 
}
\end{figure}

In figure 5   we have sketched the  Q-function
distribution functions for the output field states. We here
show this  for the squeeze parameter $r=1$, 
$\alpha_0=3$, $m=1$ and for different values of the interaction time.
We see that the Q function corresponding to the initial squeezed coherent state,  [16],
figure 5a bifurcates due to the quantum interaction between the field and the
atom. At $\lambda t=0 $ the Q function has a single-peak structure, but with
$\lambda t=T_R/2 $ we see that the Q function is composed of two well separated
components, mean while if $\lambda t=T_R $, three components have been seen figure 5c.
We show that at $t=0$ the  peak of SDFS  is observed [19].
For $t>0$ the two peaks split into two sets of counter-rotating peaks
during the collapse. At longer times the Q-function is spread out
over an angular region in the xy-plane as shown in figure 5.
  If we combine this observation with the fact that the field entropy
at this moment is almost equal to zero we can conclude that the cavity
field at $m=0 $ is in a pure state. 
On the other hand at $m=1 $
the entropy reaches its maximum and so in spite of  fact that the Q function
is composed of two parts the cavity field is in a statistical
mixture state.

\section{\bf{ Summary and concluding remarks }}

The well-known Jaynes-Cummings Model gives an exactly solvable model of a two-level 
atomic system in interaction with a radiation field. We considered the interaction with the 
 field initially is the  SDFS.   
We have been studied in this work the three  aspects, the first is
the dynamics of
the atomic inversion, the second is the field entropy,  and  later is the evolution of the output field statistics. 
The
atomic inversion have discussed and plotted against the interaction
time $\lambda t$. We found that it exhibited the conventional Rabi oscillation and collapse-revival  for the SDFS. It is dependent on the parameters of the used state as squeeze parameter and number of photons. 
We further calculated the field entropy, which is zero for a pure state and non-zero for a mixed state. In general , for the atomic radiation system the field entropy was found to be non-zero. 
We have been considered the photon number distribution of the output field.  
 We have been obtained
the Pegg-Barnett phase distribution and also plotted with some
parameters. 
  The   Q function
for some parameters  has presented 
analytically and numerically. The Q functin exhibited a variety of peaks with corresponding the photon number oscillation.  Peak separation in the Pegg-Barnett phase distribution and the Q function is associated with the onset of the collapse and revivel of the atomic population inversion.
The entropy is nearly zero when the  Q function  has exactly one peak, the greatest separation of  Q function peaks corresponds to maximum entropy. 
The effects of squeeze parameter  and photon number are  obvious from all illustrations.
The desire to realize physically certain specific quantum states such as SDFS's is under current research. It is hoped that the SDF states will find applications in the quantum nondemolition measurements and quantum optics. They may also find application in experimental situations that require low noise sensitivity.

\end{document}